\documentclass{article}

\usepackage[pdftex]{graphics}
\usepackage[polutonikogreek,english]{babel}
\usepackage[T1]{fontenc}

\usepackage{multirow}
\usepackage{url}
\usepackage{hyperref}

\usepackage{amsmath}
\usepackage{amsfonts}
\usepackage{amssymb}

\newcommand{\R}{\mathbb{R}} 			
\newcommand{\Z}{\mathbb{Z}} 			

\newcommand{\fleche}{\rightarrow}		

\newcommand{\und}[1]{_{\mathrm{#1}}}

\usepackage{graphicx}
\newcommand{\lw}{\linewidth}

\newlength{\eighty}
\newlength{\fifty}
\newlength{\ninety}
\newcommand{\InsertFigureW}[2]{\includegraphics[width=#2]{figure/#1}}

\newtheorem{questions}{Questions}

\newcommand{\BT}{
\begin{tabular}{| p{0.3\lw} | p{0.45\lw} |}
\hline
}
\newcommand{\ET}{\hline \end{tabular}}

\newcommand{\eca}[1]{{\ensuremath{\mathfrak{#1}}}}
\newcommand{\ecat}[1]{\texttt{#1}}

\newcommand{\etat}[1]{{\ensuremath{\mathtt{#1}}}}

\newcommand{\qO}{{\etat{0}}}
\newcommand{\qX}{{\etat{1}}}

\newcommand{\sr}{\alpha}

\usepackage{etoolbox}
\def\ecad#1#2{{\eca{#1}}}
\def\ecaz#1#2{{\ecat{#2}}}

\newcommand{\ECAw}[1]{%
\ifstrequal{#1}{0}{\ecad{0}{EFGH}}{}\ifstrequal{#1}{1}{\ecad{1}{AEFGH}}{}\ifstrequal{#1}{2}{\ecad{2}{BEFGH}}{}\ifstrequal{#1}{3}{\ecad{3}{ABEFGH}}{}\ifstrequal{#1}{4}{\ecad{4}{FGH}}{}\ifstrequal{#1}{5}{\ecad{5}{AFGH}}{}\ifstrequal{#1}{6}{\ecad{6}{BFGH}}{}\ifstrequal{#1}{7}{\ecad{7}{ABFGH}}{}\ifstrequal{#1}{8}{\ecad{8}{EGH}}{}\ifstrequal{#1}{9}{\ecad{9}{AEGH}}{}\ifstrequal{#1}{10}{\ecad{10}{BEGH}}{}\ifstrequal{#1}{11}{\ecad{11}{ABEGH}}{}\ifstrequal{#1}{12}{\ecad{12}{GH}}{}\ifstrequal{#1}{13}{\ecad{13}{AGH}}{}\ifstrequal{#1}{14}{\ecad{14}{BGH}}{}\ifstrequal{#1}{15}{\ecad{15}{ABGH}}{}\ifstrequal{#1}{18}{\ecad{18}{BCEFGH}}{}\ifstrequal{#1}{19}{\ecad{19}{ABCEFGH}}{}\ifstrequal{#1}{22}{\ecad{22}{BCFGH}}{}\ifstrequal{#1}{23}{\ecad{23}{ABCFGH}}{}\ifstrequal{#1}{24}{\ecad{24}{CEGH}}{}\ifstrequal{#1}{25}{\ecad{25}{ACEGH}}{}\ifstrequal{#1}{26}{\ecad{26}{BCEGH}}{}\ifstrequal{#1}{27}{\ecad{27}{ABCEGH}}{}\ifstrequal{#1}{28}{\ecad{28}{CGH}}{}\ifstrequal{#1}{29}{\ecad{29}{ACGH}}{}\ifstrequal{#1}{30}{\ecad{30}{BCGH}}{}\ifstrequal{#1}{32}{\ecad{32}{DEFGH}}{}\ifstrequal{#1}{33}{\ecad{33}{ADEFGH}}{}\ifstrequal{#1}{34}{\ecad{34}{BDEFGH}}{}\ifstrequal{#1}{35}{\ecad{35}{ABDEFGH}}{}\ifstrequal{#1}{36}{\ecad{36}{DFGH}}{}\ifstrequal{#1}{37}{\ecad{37}{ADFGH}}{}\ifstrequal{#1}{38}{\ecad{38}{BDFGH}}{}\ifstrequal{#1}{40}{\ecad{40}{DEGH}}{}\ifstrequal{#1}{41}{\ecad{41}{ADEGH}}{}\ifstrequal{#1}{42}{\ecad{42}{BDEGH}}{}\ifstrequal{#1}{43}{\ecad{43}{ABDEGH}}{}\ifstrequal{#1}{44}{\ecad{44}{DGH}}{}\ifstrequal{#1}{45}{\ecad{45}{ADGH}}{}\ifstrequal{#1}{46}{\ecad{46}{BDGH}}{}\ifstrequal{#1}{50}{\ecad{50}{BCDEFGH}}{}\ifstrequal{#1}{51}{\ecad{51}{ABCDEFGH}}{}\ifstrequal{#1}{54}{\ecad{54}{BCDFGH}}{}\ifstrequal{#1}{56}{\ecad{56}{CDEGH}}{}\ifstrequal{#1}{57}{\ecad{57}{ACDEGH}}{}\ifstrequal{#1}{58}{\ecad{58}{BCDEGH}}{}\ifstrequal{#1}{60}{\ecad{60}{CDGH}}{}\ifstrequal{#1}{62}{\ecad{62}{BCDGH}}{}\ifstrequal{#1}{72}{\ecad{72}{EH}}{}\ifstrequal{#1}{73}{\ecad{73}{AEH}}{}\ifstrequal{#1}{74}{\ecad{74}{BEH}}{}\ifstrequal{#1}{76}{\ecad{76}{H}}{}\ifstrequal{#1}{77}{\ecad{77}{AH}}{}\ifstrequal{#1}{78}{\ecad{78}{BH}}{}\ifstrequal{#1}{90}{\ecad{90}{BCEH}}{}\ifstrequal{#1}{94}{\ecad{94}{BCH}}{}\ifstrequal{#1}{104}{\ecad{104}{DEH}}{}\ifstrequal{#1}{105}{\ecad{105}{ADEH}}{}\ifstrequal{#1}{106}{\ecad{106}{BDEH}}{}\ifstrequal{#1}{108}{\ecad{108}{DH}}{}\ifstrequal{#1}{110}{\ecad{110}{BDH}}{}\ifstrequal{#1}{122}{\ecad{122}{BCDEH}}{}\ifstrequal{#1}{126}{\ecad{126}{BCDH}}{}\ifstrequal{#1}{128}{\ecad{128}{EFG}}{}\ifstrequal{#1}{130}{\ecad{130}{BEFG}}{}\ifstrequal{#1}{132}{\ecad{132}{FG}}{}\ifstrequal{#1}{134}{\ecad{134}{BFG}}{}\ifstrequal{#1}{136}{\ecad{136}{EG}}{}\ifstrequal{#1}{138}{\ecad{138}{BEG}}{}\ifstrequal{#1}{140}{\ecad{140}{G}}{}\ifstrequal{#1}{142}{\ecad{142}{BG}}{}\ifstrequal{#1}{146}{\ecad{146}{BCEFG}}{}\ifstrequal{#1}{150}{\ecad{150}{BCFG}}{}\ifstrequal{#1}{152}{\ecad{152}{CEG}}{}\ifstrequal{#1}{154}{\ecad{154}{BCEG}}{}\ifstrequal{#1}{156}{\ecad{156}{CG}}{}\ifstrequal{#1}{160}{\ecad{160}{DEFG}}{}\ifstrequal{#1}{162}{\ecad{162}{BDEFG}}{}\ifstrequal{#1}{164}{\ecad{164}{DFG}}{}\ifstrequal{#1}{168}{\ecad{168}{DEG}}{}\ifstrequal{#1}{170}{\ecad{170}{BDEG}}{}\ifstrequal{#1}{172}{\ecad{172}{DG}}{}\ifstrequal{#1}{178}{\ecad{178}{BCDEFG}}{}\ifstrequal{#1}{184}{\ecad{184}{CDEG}}{}\ifstrequal{#1}{200}{\ecad{200}{E}}{}\ifstrequal{#1}{204}{\ecad{204}{I}}{}\ifstrequal{#1}{232}{\ecad{232}{DE}}{}}
\newcommand{\ECAt}[1]{%
\ifstrequal{#1}{0}{\ecaz{0}{EFGH}}{}\ifstrequal{#1}{1}{\ecaz{1}{AEFGH}}{}\ifstrequal{#1}{2}{\ecaz{2}{BEFGH}}{}\ifstrequal{#1}{3}{\ecaz{3}{ABEFGH}}{}\ifstrequal{#1}{4}{\ecaz{4}{FGH}}{}\ifstrequal{#1}{5}{\ecaz{5}{AFGH}}{}\ifstrequal{#1}{6}{\ecaz{6}{BFGH}}{}\ifstrequal{#1}{7}{\ecaz{7}{ABFGH}}{}\ifstrequal{#1}{8}{\ecaz{8}{EGH}}{}\ifstrequal{#1}{9}{\ecaz{9}{AEGH}}{}\ifstrequal{#1}{10}{\ecaz{10}{BEGH}}{}\ifstrequal{#1}{11}{\ecaz{11}{ABEGH}}{}\ifstrequal{#1}{12}{\ecaz{12}{GH}}{}\ifstrequal{#1}{13}{\ecaz{13}{AGH}}{}\ifstrequal{#1}{14}{\ecaz{14}{BGH}}{}\ifstrequal{#1}{15}{\ecaz{15}{ABGH}}{}\ifstrequal{#1}{18}{\ecaz{18}{BCEFGH}}{}\ifstrequal{#1}{19}{\ecaz{19}{ABCEFGH}}{}\ifstrequal{#1}{22}{\ecaz{22}{BCFGH}}{}\ifstrequal{#1}{23}{\ecaz{23}{ABCFGH}}{}\ifstrequal{#1}{24}{\ecaz{24}{CEGH}}{}\ifstrequal{#1}{25}{\ecaz{25}{ACEGH}}{}\ifstrequal{#1}{26}{\ecaz{26}{BCEGH}}{}\ifstrequal{#1}{27}{\ecaz{27}{ABCEGH}}{}\ifstrequal{#1}{28}{\ecaz{28}{CGH}}{}\ifstrequal{#1}{29}{\ecaz{29}{ACGH}}{}\ifstrequal{#1}{30}{\ecaz{30}{BCGH}}{}\ifstrequal{#1}{32}{\ecaz{32}{DEFGH}}{}\ifstrequal{#1}{33}{\ecaz{33}{ADEFGH}}{}\ifstrequal{#1}{34}{\ecaz{34}{BDEFGH}}{}\ifstrequal{#1}{35}{\ecaz{35}{ABDEFGH}}{}\ifstrequal{#1}{36}{\ecaz{36}{DFGH}}{}\ifstrequal{#1}{37}{\ecaz{37}{ADFGH}}{}\ifstrequal{#1}{38}{\ecaz{38}{BDFGH}}{}\ifstrequal{#1}{40}{\ecaz{40}{DEGH}}{}\ifstrequal{#1}{41}{\ecaz{41}{ADEGH}}{}\ifstrequal{#1}{42}{\ecaz{42}{BDEGH}}{}\ifstrequal{#1}{43}{\ecaz{43}{ABDEGH}}{}\ifstrequal{#1}{44}{\ecaz{44}{DGH}}{}\ifstrequal{#1}{45}{\ecaz{45}{ADGH}}{}\ifstrequal{#1}{46}{\ecaz{46}{BDGH}}{}\ifstrequal{#1}{50}{\ecaz{50}{BCDEFGH}}{}\ifstrequal{#1}{51}{\ecaz{51}{ABCDEFGH}}{}\ifstrequal{#1}{54}{\ecaz{54}{BCDFGH}}{}\ifstrequal{#1}{56}{\ecaz{56}{CDEGH}}{}\ifstrequal{#1}{57}{\ecaz{57}{ACDEGH}}{}\ifstrequal{#1}{58}{\ecaz{58}{BCDEGH}}{}\ifstrequal{#1}{60}{\ecaz{60}{CDGH}}{}\ifstrequal{#1}{62}{\ecaz{62}{BCDGH}}{}\ifstrequal{#1}{72}{\ecaz{72}{EH}}{}\ifstrequal{#1}{73}{\ecaz{73}{AEH}}{}\ifstrequal{#1}{74}{\ecaz{74}{BEH}}{}\ifstrequal{#1}{76}{\ecaz{76}{H}}{}\ifstrequal{#1}{77}{\ecaz{77}{AH}}{}\ifstrequal{#1}{78}{\ecaz{78}{BH}}{}\ifstrequal{#1}{90}{\ecaz{90}{BCEH}}{}\ifstrequal{#1}{94}{\ecaz{94}{BCH}}{}\ifstrequal{#1}{104}{\ecaz{104}{DEH}}{}\ifstrequal{#1}{105}{\ecaz{105}{ADEH}}{}\ifstrequal{#1}{106}{\ecaz{106}{BDEH}}{}\ifstrequal{#1}{108}{\ecaz{108}{DH}}{}\ifstrequal{#1}{110}{\ecaz{110}{BDH}}{}\ifstrequal{#1}{122}{\ecaz{122}{BCDEH}}{}\ifstrequal{#1}{126}{\ecaz{126}{BCDH}}{}\ifstrequal{#1}{128}{\ecaz{128}{EFG}}{}\ifstrequal{#1}{130}{\ecaz{130}{BEFG}}{}\ifstrequal{#1}{132}{\ecaz{132}{FG}}{}\ifstrequal{#1}{134}{\ecaz{134}{BFG}}{}\ifstrequal{#1}{136}{\ecaz{136}{EG}}{}\ifstrequal{#1}{138}{\ecaz{138}{BEG}}{}\ifstrequal{#1}{140}{\ecaz{140}{G}}{}\ifstrequal{#1}{142}{\ecaz{142}{BG}}{}\ifstrequal{#1}{146}{\ecaz{146}{BCEFG}}{}\ifstrequal{#1}{150}{\ecaz{150}{BCFG}}{}\ifstrequal{#1}{152}{\ecaz{152}{CEG}}{}\ifstrequal{#1}{154}{\ecaz{154}{BCEG}}{}\ifstrequal{#1}{156}{\ecaz{156}{CG}}{}\ifstrequal{#1}{160}{\ecaz{160}{DEFG}}{}\ifstrequal{#1}{162}{\ecaz{162}{BDEFG}}{}\ifstrequal{#1}{164}{\ecaz{164}{DFG}}{}\ifstrequal{#1}{168}{\ecaz{168}{DEG}}{}\ifstrequal{#1}{170}{\ecaz{170}{BDEG}}{}\ifstrequal{#1}{172}{\ecaz{172}{DG}}{}\ifstrequal{#1}{178}{\ecaz{178}{BCDEFG}}{}\ifstrequal{#1}{184}{\ecaz{184}{CDEG}}{}\ifstrequal{#1}{200}{\ecaz{200}{E}}{}\ifstrequal{#1}{204}{\ecaz{204}{I}}{}\ifstrequal{#1}{232}{\ecaz{232}{DE}}{}}

\begin{document}

\title{ A guided tour of asynchronous cellular automata }

\author{
Nazim Fat\`es
\url{nazim.fates@loria.fr}
Inria Nancy Grand-Est, LORIA UMR 7503 \\ 
F-54 600, Villers-l\`es-Nancy, France\\
}

\maketitle

\begin{abstract}
Research on asynchronous cellular automata has received a great amount of attention these last years and has turned to a thriving field.
We present a state of the art that covers the various approaches that deal with asynchronism in cellular automata and closely related models. 
\end{abstract}

{\bf Foreword}:
This article is the preprintof an article that is to appear in the {\em Journal of cellular automata}. It is an extended verion of the invited paper that appeared in the proceedings of Automata'13, 19th International Workshop on Cellular Automata and Discrete Complex Systems, LNCS 8155, 2013, p.~15-30. 


\nocite{Hem82}
\nocite{Ach95}

\section{Introduction}

By their very simplicity, cellular automata are mathematical objects  that occupy a privileged situation in the study of complex systems. 
They are formed of a regular arrangement of simple automata, 
the {\em cells}, which can hold a finite number of states. 
Cellular automata are as mosaics with tiles that autonomously change their colour:
the cells are updated at discrete time steps and their new state is calculated according to only a local information, usually limited to the states of the neighbouring cells. 
These local laws of interaction may generate amazing behaviours at the global scale, even when they are simply expressed.

Cellular automata were initially studied by von Neumann and Ulam to study the properties of self-reproduction of living organisms with a simple ``mechanical'' tool~\cite{vN66}. Since then, they have been employed in various scientific domains. Their study can be divided into three main axes:
(1) They are dynamical systems where time, space and states are discrete. Their regular structure simplifies the mathematical definitions of the system 
but the exact or partial prediction of the trajectories of the system is often a highly challenging task.
(2) They represent a model of spatially-extended, distributed and homogeneous computing systems. 
As such, they represent an alternative to the classical computing frameworks that use sequential algorithms, variables, functions, etc.
(3) They are employed to model the numerous complex systems seen in Nature. 
Researchers have been particularly interested in the properties of self-organisation or robustness they can display.

An important feature in the definition of cellular automata regards their updating: 
in their original definition, they are updated {\em synchronously}, 
that is, all the cells change their state at the same (discrete) time step.
This global update implies a {\em strong} simultaneity: cells need to gather simultaneously the state of their neighbours, they need to process this information simultaneously, the transitions have to occur in a single time step.

Making this hypothesis of perfect synchrony has many advantages, first of all
to simplify the description of the system.
With a synchronous update, it is for instance easy to build a Turing-universal system, to ``program'' the system to obtain a given behaviour, to show that a given property is undecidable or to study under which restrictions this property becomes decidable, etc. (see the survey by Kari~\cite{Kar05} for more details). 
Synchronous updates are also a convenient tool for modelling natural or artificial phenomena: 
there is no need to take into account complex updating procedures as all the cells share the same time.

In spite of these manifest advantages, there are reasons why the hypothesis of {\em perfect synchrony} needs to be questioned:

(a)
In the context of dynamical systems, the problem is to study how cellular automaton ``react'' 
to perturbations of their updating. 
How can we interpret the potential sensitivity of the system to changes of its definition? On the contrary, what can be said if the system ``resists'' to a change of its updating scheme?

(b)
In the context of parallel computing, we ask how to design a computing device that does not require a central clock. 
Various advantages can be expected from the removal of a pace maker: increase of the speed of computations, economy of energy, simplicity of design, etc. Beyond these potential gains, developing asynchronous massively parallel algorithms represents a research challenge by itself.

(c)
When cellular automata represent a model of a natural phenomenon, 
the question is to know what triggers the transitions of the cells' state. How do we represent this source of activity in the model?
Answering is far from being simple and the argument that ``there is no global clock in Nature to synchronise the transitions'' is somewhat incomplete. 
Indeed, it can be objected that a model is a simplified representation of a phenomenon and does not need to faithfully account for all the details of ``reality''. 

All these questions raise rich problems and they are discussed in the works that we present in this survey.
The field of asynchronous cellular automata has attracted the interest of numerous authors and has evolved from a ``marginal'' to a ``respected'' topic during the last decade. The scientific production has now reached a level which makes it difficult to follow all the contributions that appear.
The purpose of this survey is thus to introduce the readers to this quite diversified ``landscape'', trying as much as possible to cover the various ``sites'' that it contains. 
As a ``guided tour'', it does not claim to be an objective description of the field:
a guided tour is by definition a circuit that takes visitors from place to place according to the arbitrary choices of the guide.
It is therefore important to bear in mind that the descriptions that will follow will be as brief as possible and should by no means prevent us from reading the texts mentioned {\em themselves}.
Our hope is that readers that are unfamiliar with cellular automata will find landmarks for their orientation and those which are interested in a particular topic will find useful references.

\section{Defining asynchronism}

Our visit begins by considering the definitions of asynchronism. 
The etymology is clear: 
\textgreek{a-sun-qr'onos} ({\em a-sun-chronos}) 
means {\em not-same-time} in Greek. 
The word thus merely indicates that there are parts of the system that do not share the same time. 
As an illustration, we may figure out a choreography where each dancer has its own pace and its own sequence of movements: the choreography may be chaotic 
but the dancers may also succeed in forming a coherent performance if some coordination is maintained between them.
 
The {\em privative} nature of the definition of a-synchronism suggests that there are many interpretations of the word. 
In fact, we are allowed to speak of asynchronism as soon as we break the framework of perfect updating. 
To date, there is no agreement on how this word should defined. 
Moreover, it is frequent that different terms are used for naming the {\em same} updating scheme. 
The definitions that we present below are thus by no means ``official'': we simply make the choice to use in priority the terms that we have employed in our own research.

\subsection{Full vs. partial asynchronism}

\begin{figure}
\InsertFigureW{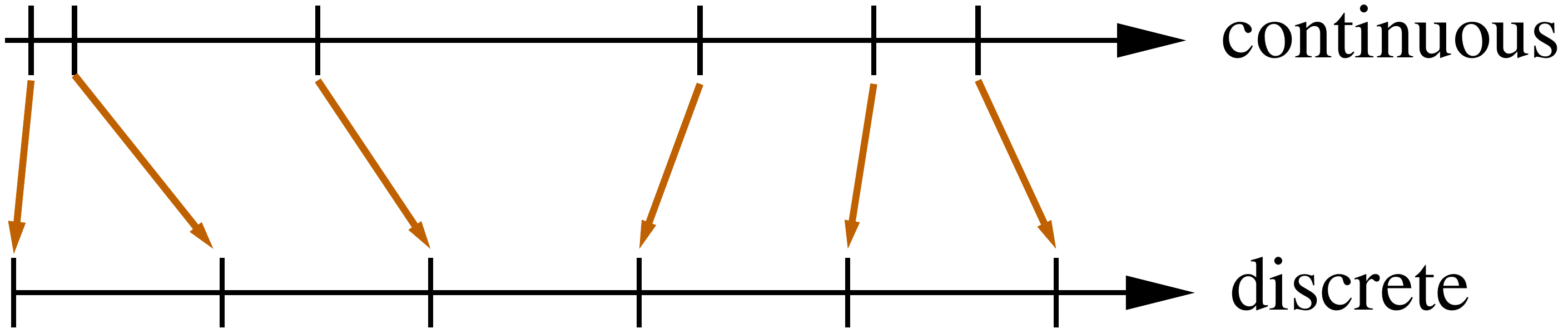}{.8\lw}
\caption{Mapping from a continuous to a discrete time scale.} 
\label{fig:rescaling}
\end{figure}

In general, asynchronism is seen as an external and uncontrolled phenomenon, 
it is thus most often modelled as a stochastic process.  The two main stochastic updating schemes that have been employed are: 
\begin{itemize}
\item
{\em fully asynchronous updating:}
At each time step, the local rule is applied to only one cell, chosen uniformly at random among the set of cells.
\item 
{\em $\alpha$-asynchronous updating:}
At each time step, each cell has a given probability $ \alpha $ to apply the rule and a probability $ 1 - \alpha $ to stay in the same state. The parameter $ \alpha $ is called {\em the synchrony rate}\footnote{ 
Note that the terms $\alpha$-{\bf a}synchronism and $\alpha$-synchronism have been used and are both relevant: $ \alpha $ can denote the name of the scheme and the synchrony rate.
We use here the term $\alpha$-{\bf a}synchronism, as it is the form that was first proposed and which has been adopted by various authors such as Regnault, Correia, Worsch, Fuk\'s, etc.
}.
\end{itemize}

Many authors consider that the fully asynchronous updating
is the most ``natural'' updating method. 
The argument that justifies this choice is a syllogism that can be decomposed as follows: (a) ``real'' time is continuous, (b) transitions occur at random moments on this continuous timeline, (c) since there is no chance of a simultaneous updating, or only a negligible chance, the only thing that matters is {\em the order} in which cells are updated, (d) this order can be obtained by a sequential stochastic sampling on the set of cells (see Fig.~\ref{fig:rescaling} and e.g. Ref.~\cite{Sch99} for a similar presentation).

It can be remarked that this argument is physically relevant if the transitions of the cellular automaton are ``infinitely'' short, that is, if the time to go from one state to another can be neglected. This is surely a valid hypothesis for some particular contexts (e.g. a radioactive disintegration) but this cannot be considered as {\em the} asynchronous updating model. 

In many cases, especially in biological systems, some synchrony between cells needs to be assumed. As this degree of synchrony is difficult to measure, the problem is not so much about choosing the ``right'' model of updating but rather to estimate the {\em robustness} of model, that is, 
if it will totally or partially resist the perturbation of its updating scheme.
In this context, the $\alpha$-asynchronous method defines a system with a continuous variation from a perfect synchronism ($\alpha = 1$) to the limiting case of full asynchronism ($\alpha \fleche 0$). 
Note however that when looking at the {\em asymptotic} behaviour of a system, a discontinuity may exist between the case $ \alpha \fleche 0 $ and the fully asynchronous case. 
Indeed, the possibility that two neighbouring cells simultaneously update their state, be it as small as wanted, may radically change the trajectory of a system.
As an example, consider the minority rule in 2D with a von Neumann neighbourhood: with a fully asynchronous updating, the two uniform fixed points are not reachable from a non-uniform configuration~\cite{FatGer09} but this is not longer true if we allow a small degree of synchrony.

It should also be noted that fully asynchronous updating is defined with a finite set of cells. The passage to the limit for an infinite set of cells needs to be done with a model that has a {\em continuous time} and the mathematical model that accurately describes a (stochastic) fully asynchronous updating is called an {\em interacting particle system}. (See e.g. Ref.~\cite{BFMM13} for examples where such systems are used for solving the density classification problem in two dimensions.)

\subsection{How to describe asynchronism?}

The other question that is generally asked when defining asynchronism is to know if the timing of the transitions should be defined with the use of a global clock or with a clock that is proper to each cell. Sch\"onfisch and de Roos call the former {\em step-driven} methods and the latter  {\em time-driven} methods~\cite{Sch99}.
It may be thought at first that ``time-driven'' methods are more adequate for making ``realistic'' simulations than ``step-driven'' methods. Indeed, it seems better to give to the cells an explicit representation of time and to avoid to artificially share a transition signal between all the cells. However, this idea needs to be examined more closely. As remarked by various authors, this distinction is somewhat artificial as it is in general possible to build a ``step-driven'' method that emulates a ``time-driven'' method, and vice versa. 
For example, as discussed before, the random updates of a fully asynchronous scheme and the updates obtained by independent clocks that use a {\em continuous} time are quasi-equivalent, up to a rescaling.
The $\alpha$-asynchronous updating can also be defined from the point of view of the cells by separating two updates by a random time which obeys a geometric law. (In other words, the probability that $ k $ time steps separate two updates of a cell is equal to: $ \alpha(1-\alpha)^{k-1} $.)

There are of course many other types of updating schemes where randomness is involved. 
For instance, one may consider {\em random sweeps} where cells are updated sequentially 
by following random permutations of the updating orders (this scheme is also called {\em random order}) or {\em fixed sweeps} where the permutation order is drawn at the beginning and kept fixed during the whole evolution of the system~\cite{Sch99}. We will also present below how to define an asynchronism which results from an imperfect transmission of the state from the neighbours (see Sec.~\ref{sec:betagamma}).

Non-random updating schemes can also be considered: for instance, in the {\em sequential ordered} scheme, cells are updated sequentially following an order that results from their spatial arrangement (for example from left to right and from top to bottom); cells can also be updated depending on their parity at even or odd time steps (see e.g.~\cite{RS97}). We refer to the work of Sch\"onfisch and de Roos~\cite{Sch99}, Cornforth et al.~\cite{CGN05}, Bandini et al.~\cite{BBV12} for the presentation of a collection of various deterministic or stochastic updating schemes.

It is also necessary to distinguish the {\em non-deterministic} schemes from the stochastic ones. 
As in classical automata theory, non-determinism means that a given subset of cells may be updated and all the possibilities are considered, regardless of their ``likeliness to appear''. The evolution of the system is thus represented by a set of configurations; this set  evolves according to the outcomes of each transition that can be applied.

The problem of the definition of asynchronism is thus completely open and
we end this section with the following questions:
\begin{questions}
What taxonomy of the updating schemes can be issued? What are the guidelines that can drive modellers for choosing a particular updating scheme? 
Under which restrictions (states, neighbourhoods, class of rules, etc.) can we establish equivalences between updating schemes?
\end{questions}

\section{Experimental approaches}

Classifying ``classical'' cellular automata has been a central theme of research and is far from being a closed question (for recent references, see e.g. the work by Schu\"le~\cite{SchSto12,SchPhd12} and the survey by Mart\'inez~\cite{Mar13}).
It can then be thought at first that classifying {\em asynchronous} rules is a daunting task because of the additional complexity that is induced by the asynchronous updating. In fact, this is only partially true as in many cases the asynchrony may ``break'' the complexity of a rule and render it more simple to study.
In this section, we discuss the contributions that qualitatively or quantitatively estimate the effects of asynchronism with numerical simulations. 

\subsection{General classifications}

In 1984, Ingerson and Buvel carried out a pioneering work where they could show that the behaviour of simple rules could be totally disrupted by simple modifications of the updating~\cite{IngBuv84}. 
Most importantly, they questioned to which extent was the behaviour of a rule the consequence of the local rule and to which extent it was due to the updating scheme.

This question was re-examined by Bersini and Detours, who explored the difference between the Game of Life and closely related asynchronous variants~\cite{BerDet94}. 
Their main observation was the existence of a ``stabilising effect'' of asynchronous updating. The experiments were made on small-size grids, no larger than 20 by 20 cells. With such lattice sizes, they were able to observe that the fully asynchronous Game of Life may ``freeze'' on some fixed-point patterns with a labyrinth-like aspect. However, more recent work has demonstrated that it was not possible to infer the large-size behaviour from these experiments and that the stabilising effect was intimately linked to finite-size effects of the numerical experiments~\cite{BloBer99,FatLif10}. 

Sch\"onfisch and de Roos gave a decisive impulse to the research on asynchronism by comparing 
various updating schemes and by exhibiting clear examples where the schemes alter significantly the behaviour of a rule~\cite{Sch99}. They gave a detailed analysis of the statistical properties of the schemes but their experiments were limited to some specific rules. 
The question thus remained open to know how these observations could be generalised to a larger class of rules.

On this basis, Fat\`es and Morvan examined how the 256 Elementary cellular automata (ECA) reacted to $\alpha$-asynchronism~\cite{FatMor05}.
To estimate the changes of behaviour of the system quantitatively, the authors used an approximation 
of the asymptotic density, that is, the value of the density that would be reached by an infinite-size system with an infinite simulation time. 
This parameter was considered as a first means to detect changes in the behaviour: a strong variation of the asymptotic density indicates that the system has undergone a transformation while an absence of variation does not necessarily imply that the system remained stable.

The 256 rules were then classified into four qualitative sets according to their responses to the variation of the synchrony rate~$ \sr$:
(a) continuous variation of the behaviour (e.g. ECA \eca{232}), 
(b) discontinuity around $ \alpha = 1 $ (e.g. ECA \eca{2} or \eca{110}),
(c) phase transition for a critical value $ \alpha\und{c} < 1 $ (e.g. ECA \eca{50}),
and (d) non-regular behaviour (e.g. ECA \eca{184}).
One of the surprising results of this study was that no direct correspondence could be drawn between 
these new classes of robustness and the previously known classes of synchronous behaviour (e.g., the informal Wolfram classes).

Similar observations were made by Bandini et al., who tested the effects of numerous asynchronous schemes on one-dimensional binary rules where the local function depends only on two neighbours (also called ``radius-1/2'' rules)~\cite{BBV12}.

\subsection{Phase transitions}
\label{sec:DP}

Blok and Bergersen were the first authors to identify the change that occurs in the Game of Life when cells are updated with a given probability~\cite{BloBer99}. 
They used $\alpha$-asynchronism to show the existence of a {\em qualitative } transition from a ``static'' behaviour, where the system would settle on fixed points, to a ``living'' behaviour, where the system evolves by forming labyrinth-like patterns that do not fixate.
The change of behaviour is a second-order phase transition, that is, the change of behaviour that separates the two qualitatively different phases obey some well-known laws from statistical physics. 
In this case, the phase transition was shown to belong to the directed percolation universality class, which means that at the critical point, the evolution of the order parameters (e.g. the density) obeys the same power laws as an oriented percolation process that serves as a reference.

Fat\`es identified that similar phenomena occurred in Elementary Cellular Automata and that no less than {\em nine} rules also displayed phase transitions. 
It was shown that the density follows a power-law decay for the critical synchrony rate, in good agreement with the behaviour expected from the directed percolation universality class~\cite{FatJCA09}. 
A unique rule, namely ECA~\eca{178}, was shown to belong to another universality class, a fact that is explained by the symmetric role that is played by $ \qO$s and $ \qX $s in the transition rule.

The phase transition occurring in the Game of life was also re-examined 
by studying how this phenomenon was affected by perturbations of the topology~\cite{FatActa10,FatLif10}. The main finding was that the critical value of the phase transition strongly depends on the regularity of the grid and that the qualitative change of behaviour becomes more difficult to observe as links between cells are removed.

Concerning other two-dimensional rules, Regnault et al. carried out a pioneering work by explaining in detail how the asynchronous minority rule displayed various types of behaviour depending on the topology on which it is applied~\cite{RegPhD08,RST09,RRT11}.
A simple puzzling observation is that the minority rule will settle out on a checkerboard or on a stripe-like pattern depending on whether the rule is defined with the von Neumann or the Moore neighbourhood. To our knowledge, there is no mathematical explanation of this empirical observation.

Remark that two different complementary views exist on phase transitions: the most common way of describing a phase transition is to establish that for an {\em infinite} system, a qualitative difference of behaviour occurs for an infinitesimal variation of the control parameter. 
An alternative approach was adopted by Regnault who could prove that for a particular rule and a {\em finite} system, the transition corresponds to a variation of the convergence time from a linear to a polynomial function of the system's size~\cite{Regn13}.

\subsection{Coalescence}
 
A curious phenomenon was remarked when comparing the evolutions of two different initial conditions that were updated with the {\em same} local rule and the {\em same} sequence of updates:
a rapid ``coalescence'' may occur, that is, the two systems take the same state and then evolve with the same trajectory (as the same sites are updated).

This phenomenon is in some cases easy to understand, as when the coalescence occurs on an attractive fixed point, but it was also observed for a non-fixed-point region of the state space (as for ECA~\eca{46}~\cite{FatMor05}). 
From a more pragmatic point of view, the following interpretation can be given: there are asynchronous systems whose evolution rapidly becomes governed by the random number generator that dictates the updates, and not by the initial condition.

Rouquier and Morvan studied systematically the coalescence phenomenon for the 256 ECA~\cite{RouPhD08, RouMor09}. 
Their study revealed that it was possible to observe that some ECA would always coalesce, while others would never coalesce, and that there existed some rules which displayed a phase transition between a coalescing and non-coalescing behaviour. 
It is an open problem to explain analytically the non-trivial cases of rapid coalescence. It is also interesting to compare these results with those obtained by other methods of coupling (see e.g.~\cite{SanLop06,Rou08}).

\subsection{Asynchronous information transmission}
\label{sec:betagamma}

While the approaches of asynchronism studied so far are based on the dichotomy updated versus not updated, Bour\'e et al. defined a model of asynchrony which considers imperfect communications between neighbours~\cite{Bour12,BouPhD13}.  
This approach is declined in two versions, called $\beta$- and $ \gamma$-asynchronism, which respectively consider stochastic failures of the communication of a state to the whole neighbourhood or to each neighbour independently.

Among the various observations made with these two types of asynchronism, the most intriguing  
phenomenon is the disappearance of some, but not all, of the phase transitions that were obtained with the $\alpha$-asynchronism. 
More precisely, ECA \eca{6}, \eca{38} and \eca{134} do not show any transition for $\beta$- and $ \gamma$-synchronism.
ECA \eca{58} gives an even more puzzling case as it does show a phase transition for  $\alpha$- and $\beta$-asynchronism but {\em not} for $\gamma$-asynchronism. It is an open problem to understand the origin of such radical differences of responses to the rate of transmission failures.

Experiments also displayed that in some cases, quantities can be conserved when using only a particular model of asynchronism (e.g., ECA \eca{50} has some parity conservation with $\beta$-asynchronism but not with  $\alpha$-asynchronism). 
This underlines the necessity to continue to ``invent'' various perturbations of the classical updating in order to gain insight on how cellular automata are dependent on their updating schemes.

\subsection{Other variants}

An interesting development on the work of asynchronism concerns how it mixes with traditional noise, that is, on randomness imposed on the {\em state} of the cells that compose the automaton.
An early reference that addresses this question is given by Gharavi and Anantharam, who revisited a well-known result of Toom and who considered delays in the cells' communications~\cite{GhaAna92}. 
We refer to the work of Kanada~\cite{Kan94} on the 256 ECA rules, and to the work of Mamei et al.~\cite{MRZ05} for additional insights into this problem of mixing noise and asynchronism. 

More recently, Silva and Correia gave a detailed account on how some ECA can react to asynchronism combined with noise~\cite{SilCor13}. 
Interestingly, they propose to evaluate the robustness according to the difference patterns. This brings them to introduce a sampling compensation in order to cope with less frequent updates. 

The case of asynchronous models simulated on a non-regular topology was tackled by Baetens et al., who examined an asynchronous updating with a non-regular topology generated with a Voronoi tessellation~\cite{BWB12}.

\bigskip
To conclude this section, it seems that only a small part of the universe of asynchronous cellular automata has been 
explored so far. This brings us to put an emphasis on the following questions:
\begin{questions}
What is a good protocol to numerically estimate the changes of behaviour induced by asynchrony? 
What are the relevant order parameters to quantify these changes? 
How common is it to observe discontinuities of behaviour induced by a {\em continuous} change of the updating scheme?
\end{questions}


\section{Analytical approaches}

We now turn our attention to the mathematical analysis of asynchronous cellular automata. 
It is important to remark that although this part is presented separated from the previous one, there is a {\em joint} movement of going from simulations to analysis and back. (This co-development is not necessarily done by the same authors of course.)

\subsection{Markov chain analysis and classifications}

Agapie et al. conducted one of the first analytical studies of asynchronous rules using Markov chain theory~\cite{AgaHM04}. 
They focused on several models of finite cellular automata with fully asynchronous updating. However, as far as we could understand, their analysis was limited to a specific case where the local rule was stochastic, totalistic, symmetric with respect to an exchange of \qO s and \qX s, and with positive rates (the probability to reach each state is strictly positive). 
It is worth noting that the number of borders of a configuration is a central parameter in their analysis and that this parameter is also found in various other approaches.

One of the first analytical results of classification were given by Fat\`es et al. who analysed the doubly-quiescent ECA~\cite{FMST06}. 
In this study, the 64 rules considered are classified according to their worst expected convergence time to a fixed point. This time falls in the following classes: logarithmic time, linear time, quadratic time, exponential time and non-converging rules
\footnote{The classes are here given with a ``rescaled time scale'' where one step corresponds to as many random updates as there are cells in the finite ring.}. 
The visual inspection of the space-time diagrams of the rules of each class shows a good correspondence between the visual ``behaviour'' and the class. In other words, the time of convergence to a fixed point is not an ad hoc parameter 
but does capture a part of the ``behaviour'' of the stochastic rules.

These results were later partially extended to the more difficult case of $\alpha$-asynchronous updating by Regnault et al.~\cite{Reg06}, while Chassaing and Gerin examined the continuous limit of the processes when the grid was made infinite~\cite{ChaGer07}.

Fat\`es and Gerin also examined how to classify the two-dimensional totalistic 
rules with fully asynchronous updating~\cite{FatGer09}. They proposed a partial classification of 64 rules and an analysis of the convergence of some well-known rules.
Among the interesting phenomena remarked, they exhibited a list of rules which showed an ``erratic'' behaviour: the question was to determine if these rules were exhibiting a non-converging behaviour or a ``metastable'' behaviour, that is, if a (long) random sequence of updates could drive the system 
to a fixed point. 
By adapting techniques from automatic planning, Hoffmann et al. could solve this problem for a specific rule and showed that it converged to a fixed point in (at most) exponential time~\cite{HFP10}.

Readers interested in the classification of rules with regard to their convergence time can refer to a recent synthesis note~\cite{FatesAuto13b} and a recent work on the fast convergence of the ECA rules~\cite{FatFast14}.

\subsection{Detailed analysis of the asymptotic densities}

As mentioned above, for the $\alpha$-asynchronous systems, the study of the asymptotic density was mainly made with numerical simulations.
By focusing their efforts on eight simple ECA rules, Fuk\'s and Skelton succeeded to give an exact calculation of this density~\cite{FukSke11}. 
They considered infinite systems where the initial condition was generated by a Bernoulli measure and determined how the asymptotic density varies as a function of the initial density (that is, the parameter of the Bernoulli measure). 
Such exact results are generally rather difficult to obtain and it is an open problem to extend them to a wider class of rules.

Following this direction of research, Fuk\'s and Fat\`es considered a development of Gutowitz's ``local structure theory'': 
contrary to a classical mean-field approach where the state between neighbouring cells is assumed to be uncorrelated, correlations of order 2 or larger are taken into account to try to predict the asymptotic density of the system~\cite{FukFat14}. It was shown that this approach does detect the occurrence of phase transitions. The limit is that the position of the critical synchrony rate remains difficult to find: for some rules, even approximations with nine cells cannot predict precisely the position of the critical threshold that separates the active and inactive phases.


\subsection{Reversibility}

As mentioned above, the asynchronous updating of a system does not perturb its fixed points. However, when the updating is stochastic cycles no longer exist and one needs to re-examine the meaning of reversibility.
One such interpretation was proposed by Das et al., 
who define reversibility as a possibility to return to the initial condition in the case where the updating sequence (or ``update pattern'') could be set freely. 
They studied which are the Elementary Cellular Automata with null or periodic boundary conditions that allow to obtain such a form of ``cycles''~\cite{SMD12,DSS12}.
 
Another point of view considered the case of fully asynchronous updating: as the evolution of the system is adequately described by a Markov chain, reversibility is identified with the property of recurrence of this chain~\cite{SFD14}. A classification of the ECA rules into three classes was then proposed based on this tool: 
(a) The {\em recurrent} rules are those which make the system always return to its initial condition. 
(b) The {\em irreversible} rules are those which contain initial conditions which are never returned to after a (random) time. Among this class, 
(c) the {\em strongly irreversible} rules are those which contain a state that is never returned to as soon as it is updated. 
It is an open problem to determine how to extend these results to a wider class of systems, in particular to deal with the case of infinite-size systems.

Wacker and Worsch also examined the question of reversibility of asynchronous cellular automata~\cite{WaWo12}. In their work, a rule is said to be reversible if there is another rule whose state-transition graph is the ``inverse'' of the original. The novelty with respect to the synchronous case is that the out-degree of the nodes is no longer equal to one as a single configuration can lead to many others. 
Interestingly, the results presented on ECA are not far from those found in Ref.~\cite{SFD14} and it is an open question to determine which are the conditions that make the two points of view equivalent.

\subsection{m-asynchronous models and their topological properties}

The study of the dynamical properties of cellular automata, such as injectivity, surjectivity, permutivity, etc., has been
a central topic in the theoretical considerations of the field (see e.g.~\cite{Kar05}). 
Manzoni examined how these properties could be re-defined and studied in the asynchronous updating context~\cite{Man12}.  

This work was taken a step further by Dennunzio et al., who developed the notion of {\em m-asynchronous} cellular automata 
in order to generalise the various updating methods used so far~\cite{DFMM13}. They provided a formal framework to describe the updating probabilities on each cell, even in the case where the size of the system is infinite, and produced various theorems that allow to deal with the non-deterministic nature of the updating.

For more details on this line of research, we refer to the recent survey by Formenti where more details and examples can be found~\cite{For13}.

\bigskip
To synthesise, the contributions met in this section show the necessity to adapt the tools to the stochastic process theory for the specific case of cellular automata. This brings us to ask:
\begin{questions}
What is the position of asynchronous cellular automata with respect to stochastic cellular automata? (a mere subset?) 
What are the analytical tools that can ease the analysis of the Markovian systems obtained with random updates? 
\end{questions}

\section{Computing with asynchronous cellular automata}
We now consider the contributions related to the computing abilities of asynchronous models and briefly describe the techniques that have been proposed to construct such (virtual or real) computing objects.

\subsection{Simulation of (a)synchronous models by (a)synchronous models}

Nakamura was among the first authors to investigate how to compute
with an asynchronous cellular automaton~\cite{Naka74,Nak81}.  
He described several techniques to construct a universal rule and showed how to simulate a given $ q $-state deterministic rule with an asynchronous rule that has the same neighbourhood and whose state space is extended to $3q^2$ states 
(see also Lipton et al.~\cite{LipMS77}, Toffoli~\cite{Tof78} and Nehaniv~\cite{Neh04} for similar constructions).
The construction relies on the idea that when a cell is updated, it then waits the neighbouring cells to ``catch up'' and makes the next transition only when all its neighbours are up to date. Additionally, it keeps its old state available for the neighbouring cells in order for them to perform the ``right'' transitions. This construction was later improved by the use of only $ q^2 + 2q$ states by Lee, Peper et al.~\cite{LAPM04,PAL10}.

Peper et al. also proposed to consider the case where a cell can ``activate'' their neighbouring cells and showed that the cost in the number of states for the simulation of $q$-state rule could be reduced to $ \mathcal{O}(q\sqrt{q}) $ states~\cite{PIKM02}. 

Other discussions on the universality of asynchronous rules are found in the study by Takada et al., in which many important arguments and useful references can be found~\cite{Tak06}. 
In particular, the authors present a result showing the existence of an asynchronous, rotation-symmetric rule with 15 states and von Neumann neighbourhood that has the property of universal construction and computation.

An alternative point of view was given by Golze who simulates an $n$-dimensional synchronous rule with an asynchronous rule defined on a space with $ n+1$ dimensions~\cite{Gol78}. 
This solution simplifies the problem as there is no longer the need to save the previous and the current state in order to achieve correct computations.
Another advantage of having an additional dimension is to read one state of the synchronous simulated system (guest) on the asynchronous simulating system (host): it simply corresponds to reading a line (or a hyperplane) of the host. 
This technique, called ``global synchronisation'', is presented as a means to solve various problems, such as the Firing Squad Synchronisation Problem, which would not be solvable without this requirement. 
However, it can be noted that this technique can be interpreted as the ``deployment'' of Nakamura's technique on an additional dimension. Reciprocally, one can also see Nakamura's technique as the ``compressed'' version of Golze's solution, where only the necessary information is retained. 

The case where asynchronous computations have to be made with stochastic {\em and} asynchronous components was tackled by Wang~\cite{Wan91}. Unfortunately, this author does not position his work with regard to the previous contributions (Nakamura, Golze) and it is difficult to see if this proposition significantly differs from the previous achievements.

An original way to simulate a universal Turing machine with a fully asynchronous updating has also been proposed by Dennunzio et al.~\cite{DFM12}. 
The authors introduce the notion of ``scattered strict simulation'' in which they tolerate that only a subset of cells is used to perform the simulation. They find that asynchrony induces a quadratic slowdown compared to the speed of the simulated Turing machine.

\subsection{Computations and order-independence}

A key observation in the theory of asynchronous systems relies in the property of what we could call ``non-overlapping influences'': if two cells $ c $ and $ c' $ are such that the neighbourhood of $ c $ and $ c' $ do not overlap (that is, have no cell in common), it does not matter whether $ c $ is updated before $ c' $, or $ c' $ before $ c $,  or both of them are updated at the same time. The study of this property has given birth to various works that we now examine.

G\'acs was one of the first authors to determine if the evolution of an asynchronous system could be independent of the order of updating~\cite{Gac01}. He showed that although this property was undecidable, there exists a sufficient condition to verify this independence.

This question was later re-examined by Mortveit, Macauley et al., who studied in which cases repetitions of sequential updates on Elementary Cellular Automata (ECA) could produce a set of periodic points that would be independent of the updating order~\cite{MM08,MacMor10,McAul11}. 
This conducted the authors to present a list of 104 ECA which display such an update independence. 
Their work also uses an original representation of ECA that differs from the classical Wolfram code and that could prove useful for future analysis of asynchronous systems. 
(Another notation is presented in Ref.~\cite{FatesPhD,FMST06}).

Order-independence was also a key point considered by Worsch, who examined how to simulate an arbitrary rule by a universal asynchronous simulator~\cite{Wor13}. 
He extended Golze's results by tackling a large scope of updating policies: {\em purely asynchronous} (no restriction on the set of cells to update), $\alpha$-asynchronous, {\em N-independent} (where two neighbouring cells are never updated at the same time), and non-deterministic fully asynchronous. He showed that for each such policy, there is a universal rule (the host) that can ``simulate'', in a particular sense, any other guest. Worsch's work raises many questions, in particular as to how to properly define the notion of simulation of an asynchronous rule by another. (See Ref.~\cite{AST13} for some reflexions made in the context of stochastic cellular automata.)

We also point out that Vielhaber has designed a formal framework in which the computations of functions on finite binary rings ($\Z/n\Z$) are made not by changing the local rule but by a proper use of  the order of updating on a {\em fixed} rule~\cite{Viel13}. 
In particular, he showed that ECA \eca{57} with periodic boundary conditions was a rule especially adapted for such a purpose. 
Interestingly, this technique could be generalised to make this particular rule Turing-universal in the sense that the computation of an algorithm could be done only by setting up the proper sequence of updates.

\subsection{Models of concurrency }

Among the early references that can be found on asynchronous cellular automata, Priese wrote a note where he considers (two-dimensional) cellular automata as a particular case of asynchronous rewriting systems (called Thue-systems) and widens the scope by considering also the case where more than one cell may be re-written at a time (the overlapping problem)~\cite{Pri78}.  He uses his construction to show how to build asynchronous circuits which are equivalent to asynchronous concurrent Petri nets. 

Following this path, Zielonka examined how asynchronous rules could be used to describe the situations of concurrency that arise in distributed systems~\cite{Ziel93}. 
Pighizzini clarified the computing abilities of Zielonska's models~\cite{Pig94} and the problem of how to turn non-deterministic B\"uchi asynchronous cellular automata into deterministic models was solved by Muscholl~\cite{Mus94}. 
Droste generalised to partially ordered multisets (pomsets) the original notion of Zielonska's asynchronous mappings~\cite{Dro96}; these questions were later re-investigated by Kuske~\cite{DGK00,Kus00,Kus07}. 

With similar preoccupations, Hagiya et al. used formal methods from logic to verify the properties of some rewriting systems, showing the links between their approach and (a)synchronous systems~\cite{HTYS04}.

\subsection{Asynchronous circuits}

Another major field of research on asynchronous cellular automata was developed by Peper, Lee and their collaborators.
In their constructions, asynchronous computations are realised with particles that follow Brownian movements and which interact through special ``gates''~\cite{APL04,APL04b,LAPM05,PLI10}. 
These constructions result in delay-insensitive circuits that are Turing-universal (see e.g.~\cite{Lee11,LeeZhu12} and references therein). 

Recently, Schneider and Worsch presented a 3-state rule that uses Moore neighbourhood which can simulate any delay-insensitive circuit~\cite{SchWor12} 
and
Lee et al. presented a generalisation of their work in the context of number-conserving cellular automata~\cite{LeeIZ12}.

\bigskip

To end this section, we propose to put an emphasis on the following questions:

\begin{questions}
What is a good definition of the simulation of an (a)synchronous system by another (a)synchronous system? What are the techniques to simulate various asynchronous systems by other asynchronous systems? (e.g.: When can an $ \alpha$-asynchronous system with a given synchrony rate simulate another system with a different synchrony rate?)
\end{questions}

\section{Modelling with asynchronous cellular automata}

Asynchronous rules have been designed for specific goals such as finding new algorithms, developing new types of computing devices, modelling various natural or artificial complex systems, etc. In fact, giving a representative view of these contributions would necessitate a whole independent survey. 
The task is all the more difficult as often authors use asynchronous updating without even mentioning it.
For the sake of brevity, we will thus only give a few entry points, concentrating on the papers where the question of the updating is explicitly discussed.

\subsection{Game theory and Ecology}

As mentioned earlier, the hypothesis of perfect synchrony poses the problem of the  {\em realism} of a model: How to interpret the behaviours that are only due to the updating and not to the rule that governs the cells? 
Huberman and Glance gave evidence of the existence of such ``artifacts'' and 
challenged the validity of the simulations of spatially-extended models of the Prisoner's dilemma: a change in the updating models brings out new conclusions, drastically opposed to what was known with the classical models~\cite{HubGla93}.
This question was re-examined by Newth and Cornforth who showed that asynchronism could also lead to the observation of new cooperative phenomena not seen in the synchronous setting~\cite{NewCor09}. (See also Ref.~\cite{New09} for a non-spatially-extended version of the problem.)

Grilo and Correia also considered this problem but instead of restricting their study to the fully asynchronous scheme, they employed $\alpha$-asynchronous updating to explore a wide range of degrees of synchrony~\cite{GriCor11}. Their study revealed that the changes induced by smooth variations of the synchrony rate may brutally affect the level of cooperation in the system, a behaviour that is strongly reminiscent of the second-order phase transition seen in binary systems (see Sec.~\ref{sec:DP}).
Saif and Gade also investigated this issue and found that there was a first order transition between a regime with an all-defector state to a mixed state~\cite{SaiGad09}. All these works share in common the conclusion that some previously observed equilibrium states are artifacts of a synchronous updating on a regular lattice.

Ruxton and Saravia have discussed the importance of the ordering in the context of ecological modelling, studying a stochastic model of colonisation of an environment by a species~\cite{RuxSar98}. They argue in favour of adapting the updating scheme to the physical reality of the system that is modelled.  The authors also emphasise the need to describe precisely the updating scheme that is used to facilitate the reproducibility of the experiments. These arguments come to strengthen the need for studying in detail the ``emergence phenomena'' that are seen in Ecology and question whether the predictions of the models can be observed in ``real-life'' systems~\cite{RohLGR97,CarEtal08}.

\subsection{Synchronisation in physical and biological systems}

In an approach close to the work of Turing on morphogenesis~\cite{Tur52}, 
Gunji used asynchronous cellular automata to analyse the pattern formation mechanisms that occur in molluscs~\cite{Gun90}. 
Another interesting biological example is given by Messinger et al., who investigated the link between emergence of synchrony and the simultaneous opening and closing stomatal arrays in plants~\cite{MMP07}.

In Physics, we mention the work of Le Ca\"er~\cite{LeC89} and Radicchi et al.~\cite{RVMO07}, who studied how numerical simulations of cellular systems would be dependent on the updating. In the latter work, the local rule is itself stochastic; the authors emphasise the fact that neither totally synchronous nor totally asynchronous updating is fully relevant for modelling natural systems.

\subsection{Problem solving}

Tomassini and Venzi~\cite{TomVen02}, Capcarrere~\cite{Cap02} and Nehaniv~\cite{Neh03} have studied how asynchronous rules solve the density classification problem and the global synchronisation problem. 
Readers interested in this issue are referred to a study by Vanneschi and Mauri, in which an enlightening discussion on these various contributions is found and where the authors present findings of robust and generic rules~\cite{VanMau12}.

Suzudo examined the use of genetic algorithms to find mass-conservative (also called number-conserving) asynchronous models that would generate non-trivial patterns~\cite{Suz04,Suz04b}. 
He classified these patterns into three categories: checkerboards, stripes and sand-like. 
In this work asynchronism is mainly used to ensure that number of particles remain constant, but it is also a useful technique for generating regular patterns out of randomness: this task is known to be difficult in the synchronous setting (see e.g.~\cite{FatDCP12}).

Beigy and Meybodi investigated how asynchronous systems perform learning tasks and presented applications of their work for pattern generation and control of cellular mobile networks~\cite{BeiMey08}. 

It is also worth mentioning that Lee et al.~\cite{LAP07} and Huang et al.~\cite{HLSP13} designed models of self-reproduction that use asynchronous models (also called self-timed cellular automata).

\subsection{Other problems}

On the simulation side, Overeinder and Sloot were among the first to examine how to deal with the simulation of asynchronous automata on distributed systems~\cite{OveSlo93}.
Bandman and other authors studied how to simulate chemical systems with asynchronous cellular automata~\cite{Ban10,ShaElo11}.
Hoseini et al. made an implementation of asynchronous rules with FPGAs~\cite{Hos10}. They propose a particular design of the FPGA in order to construct a ``conformal computer'', that is, a computer made of physical cells ``arrayed on large thin flexible substrates or sheets. Sheets may be cut, joined, bent, and stacked to conform to the physical and computational needs of an application''. 

Original applications were considered by Bandini et al., who used asynchronous rules with memory for the design of an illumination facility~\cite{BBVA09} and by Minoofam et al., where asynchronism produce calligraphic patterns in the Arabic Kufic style~\cite{Minoo10}. (Unfortunately, this paper lacks precision on the model that is used). 

\bigskip
As we have seen in this section, there is a broad range of domains where asynchronous models have been employed and those which we cited above are only a small part. 

\begin{questions}
How can we develop a unified simulation environment to facilitate the comparison of various updating schemes? Is there a method for identifying the artifacts that are due to a perfect synchronous updating? Are such effects avoidable?
\end{questions}

\section{Asynchronism in other discrete models of complex systems}

We end this guided tour on an opening on the use of asynchrony in the systems whose structure is close to cellular automata.
Again, this is such a wide topic that we will indicate only a few entry points to the literature.

\subsection{Links with multi-agent systems}

One first proposition to link the updating in multi-agent systems and cellular automata was made by Cornforth et al., but the models they studied are in fact standard asynchronous cellular automata~\cite{CGN05}.
 Spicher et al. considered the question of how to ``translate'' a multi-agent system with sequential updating into a synchronous cellular automaton~\cite{SpiFatSim10}. 
So-called {\em transactional cellular automata} were defined to model the movements of particles between neighbouring cells. One positive effect of using a synchronous cellular automaton 
is to remove the spurious effects that could be linked to a particular updating order. (The authors give the example of diffusion-limited aggregation.)

The link between large-scale multi-agent systems and asynchronous cellular automata was also examined by To{\v{s}}i{\'c}~\cite{Tos11}. 
This author argues that the structure of cellular automata needs to be modified in several aspects, among which it should be made asynchronous, in order to serve as a basis for modelling large groups of interacting agents.

An alternative approach to model (discrete) multi-agent systems was proposed by Chevrier and Fat\`es, who studied the dynamics of a simple multi-turmite systems, also known as multiple Langton's ants. 
Their formalism, inspired by cellular automata, captures the possibility to have {\em synchronous} interacting agents~\cite{CheFat10}. The difficulty relies in describing how to solve conflicts that occur when two or more agents simultaneously want to modify the environment. The solution relies on a framework invented by Ferber and M{\"u}ller called {\em influence-reaction}~\cite{FerMul96}.
Belgacem and Fat\`es later extended this work by considering a wider range of updating procedures and discovered some phenomena (e.g., gliders) that resisted variations in the updating choices~\cite{BelFat12}.

Interesting observations were also made by {\c{S}}amilo{\u{g}}lu et al. who analysed the clustering effects in a group of self-propelled particles~\cite{SGK06}. 
They model asynchronism with the introduction of delays in the updating and observe that the coherence of the groups are strongly diminished as the bounds on the delays are increased.

\subsection{Lattice-gas cellular automata}

Lattice-Gas Cellular Automata (LGCA) can be seen as a ``bridge'' between cell-based updating and agent-based updating. Applying asynchrony in this context is not a straightforward operation and a first proposition of an asynchronous LGCA was made by Bour\'e et al.~\cite{BFC13}. 
In their model, movements of particles are defined explicitly, like in multi-agents, but the updating is made cell by cell, like in classical cellular automata. 
Various responses to asynchrony are observed depending on the patterns on which the system stabilises. 
In particular, strange patterns such as checkerboards are shown to disappear where randomness in the updating is added. It is an open problem to know if an infinitesimal amount of asynchrony is sufficient to destroy this pattern.

These first results show the need to explore various possibilities to define an asynchronous LGCA. 
In particular, it is interesting to look at a way to update particles independently.

\subsection{Automata networks, neural networks and other models}

The effect of asynchronous updating in genetic regulatory networks has also been investigated by many authors. 
Aracena et al. introduce a labelled directed graph that allows to determine to which extent deterministic update schedules are equivalent~\cite{AGMS09}.
Demongeot et al.~\cite{Sene10}, and Noual~\cite{Nou11} examine the robustness of the system under the variation of updating schemes and this perturbation is coupled with various topological modifications of the network such as adding or removing links in the graph or changing boundary conditions.

The question of the effect of the updating in neural networks has been discussed by Scherrer~\cite{Sch05}, Taouali et al.~\cite{Taouali11}. In particular, the latter authors introduce an interesting distinction between the use of (a)synchronous updating at the modelling level and at the implementation level.

In the context of ``amorphous computing'', Stark discussed the computing abilities of a computing medium formed out of non-regularly placed cells which obey asynchronous updating~\cite{Sta00,Sta13}. This author suggests that asynchrony plays an enhancing role for the computing abilities of such systems.

We mention that the differences between synchronous and asynchronous updating were also investigated in coupled map lattices~\cite{LumNic94,RBJ98,AbrZan98}. 
Similarly, the effects of the updating in the Asymmetric Exclusion Process (ASEP) have been studied by Rajewsky et al.~\cite{RSS98}. 
Tomita et al. studied asynchronous graph-rewriting systems and showed how to make such systems simulate their synchronous counterparts~\cite{TMK07}.

\bigskip
To end this section, we wish to highlight the following questions:
\begin{questions}
What light can be shed by asynchronous cellular automata on other closely related models and vice versa? Can we transfer the techniques used to analyse the simple asynchronous cellular systems to more complex models? 
What is the interplay between the regular topology of cellular automata 
and the regularity of their updating?
\end{questions}

\section{Closing words}

This guided tour allowed us to consider the various contributions that deal with the question of asynchronism in cellular automata and closely related models. 
As we have seen, asynchrony is a {\em privative} property that does not in itself specify a system: 
there are plenty of ways to construct an asynchronous system and all of them are a priori valid. 

One of the main current challenges is to continue to explore this question with a {\em joint work} of mathematical  analysis and numerical simulations. 
As we have seen, analytical results have been more difficult to obtain than numerical ones, but the situation is progressively changing as more techniques from the probability theory are being developed for the specific case of cellular automata.
We find it rather amazing that it is still an open question to determine the convergence time of some simple binary rules~\cite{FatesAuto13b,FatFast14}.

It is also important to clarify the position of asynchronous cellular automata into the wider field of stochastic cellular automata. 
Indeed, asynchrony is not a mere type of noise: recall for example that the addition of asynchrony to a deterministic model does not change its fixed points. However, many phenomena such as the existence of singularities or phase transitions can certainly find their explanations using the stochastic process theory and statistical physics. 

As far as modelling is concerned, the main challenge would be to carry out an experimental work to validate some models of asynchronism or to dismiss some others for {\em specific situations}. As we mentioned earlier, the no-global-clock argument --- ``Nature does not possess a clock to synchronise the transitions.'' --- cannot be received directly and be taken alone as a valid objection to the use of synchronous models. Instead, we consider that studying a {\em single} updating scheme is not sufficient and one should instead compare various possibilities to model a ``natural computing'' system. 

The principal observation from this guided tour is the existence of a great variety of approaches to asynchronism.  
This raises the question of what {\em is} time in the context of computer science and numerical simulations. 
The positive sciences define time as an object -- identified with $ \R$, with $ \Z$, a collection of coordinates, etc. -- but it may well be that time is not some ``thing'' that can be studied ``objectively''.

Does this mean that time is subjective and that our models should reflect this subjectivity? 
Such considerations would lead us out of the scientific method and would therefore be dismissed as non rational. 
Can we then escape the dilemma of ``objective versus subjective time''? 
No simple answer can be given and for sure, time is 
one of the central problems of philosophy.
It is certainly not a coincidence if one the most important philosophical contributions of the past century bears as title: {\em Sein und Zeit} ({\em Being and Time}).

\section*{Acknowledgements}

May all the persons that have contributed to the elaboration of this text receive our sincere expression of gratefulness.  
The author knows how much he owes to them, including the organisers of Automata'13 and their precious support, and the anonymous referees, whose constructive propositions were greatly appreciated. In echo to Heraclitus' word that ``the invisible harmony is greater than the apparent one'' (DK B54), he hopes that they will pardon him for not being named here. 
The author is aware of the {\em limits} of this text and he will be grateful to all the corrections, indications, remarks, and suggestions that will be given to him.

\bibliographystyle{plainurl}
\bibliography{bib-asynch,bib-Fates,bib-CA}


\end{document}